\documentclass[aps,prl,groupedaddress,superscriptaddress,twocolumn,showpacs,floatfix]{revtex4-1}
\usepackage{graphicx}
\usepackage{times}
\usepackage{nicefrac}
\usepackage{amsmath}
\usepackage{amsfonts}
\usepackage{amssymb}
\usepackage{amsthm}
\usepackage{epsf}
\usepackage{bm}

\usepackage{dcolumn}
\newcolumntype{.}{D{x}{}{-1}}

\newcommand{\aZ}{\alpha Z}
\begin{document}

\title{\emph{g} Factor of Lithiumlike Silicon:\\
New Challenge to Bound-State QED
}
%
\author{D.~A.~Glazov}
\affiliation{Department of Physics, St. Petersburg State University, Universitetskaya 7/9, 199034 St.~Petersburg, Russia}
\author{F.~K\"ohler-Langes}
\affiliation{Max-Planck-Institut f\"ur Kernphysik, Saupfercheckweg 1, D-69117 Heidelberg, Germany}
\author{A.~V.~Volotka}
\affiliation{Department of Physics, St. Petersburg State University, Universitetskaya 7/9, 199034 St.~Petersburg, Russia}
\affiliation{Helmholtz-Institut Jena, Fr\"obelstieg 3, D-07743 Jena, Germany}
\affiliation{GSI Helmholtzzentrum f\"ur Schwerionenforschung GmbH, Planckstra\ss e 1, D-64291 Darmstadt, Germany}
\author{F.~Hei\ss e}
\affiliation{Max-Planck-Institut f\"ur Kernphysik, Saupfercheckweg 1, D-69117 Heidelberg, Germany}
\affiliation{GSI Helmholtzzentrum f\"ur Schwerionenforschung GmbH, Planckstra\ss e 1, D-64291 Darmstadt, Germany}
\author{K.~Blaum}
\affiliation{Max-Planck-Institut f\"ur Kernphysik, Saupfercheckweg 1, D-69117 Heidelberg, Germany}
\author{G.~Plunien}
\affiliation{Institut f\"ur Theoretische Physik, Technische Universit\"at Dresden, Mommsenstra{\ss}e 13, D-01062 Dresden, Germany}
\author{W.~Quint}
\affiliation{GSI Helmholtzzentrum f\"ur Schwerionenforschung GmbH, Planckstra\ss e 1, D-64291 Darmstadt, Germany}
\author{V.~M.~Shabaev}
\affiliation{Department of Physics, St. Petersburg State University, Universitetskaya 7/9, 199034 St.~Petersburg, Russia}
\author{S.~Sturm}
\affiliation{Max-Planck-Institut f\"ur Kernphysik, Saupfercheckweg 1, D-69117 Heidelberg, Germany}
\author{G.~Werth}
\affiliation{Institut f\"ur Physik, Johannes Gutenberg-Universit\"at, D-55099 Mainz, Germany}
\begin{abstract}
The recently established agreement between experiment and theory for the \emph{g} factors of lithiumlike silicon and calcium ions manifests the most stringent test of the many-electron bound-state quantum electrodynamics (QED) effects in the presence of a magnetic field. In this Letter, we present a significant simultaneous improvement of both theoretical $g_\text{th} = 2.000\,889\,894\,4\,(34)$ and experimental $g_\text{exp} = 2.000\,889\,888\,45\,(14)$ values of the \emph{g} factor of lithiumlike silicon $^{28}$Si$^{11+}$. The theoretical precision now is limited by the many-electron two-loop contributions of the bound-state QED. The experimental value is accurate enough to test these contributions on a few percent level.
\end{abstract}


\maketitle
%
%
\textit{Introduction.} ---
The magnetic moment of elementary particles and simple systems is a perfect tool for testing fundamental theories. High-precision \emph{g}-factor measurements in highly charged ions \cite{haeffner:00:prl,verdu:04:prl,sturm:11:prl,sturm:13:pra,wagner:13:prl,koehler:16:nc,sturm:17:a} in combination with elaborate theoretical investigations (see, e.g., \cite{shabaev:15:jpcrd,harman:18:jpcs} for reviews) have provided the most stringent test of bound-state QED in the presence of a magnetic field up-to-date. Moreover, these studies resulted in the most accurate value of the electron mass~\cite{sturm:14:n,CODATA14,zatorski:17:pra,czarnecki:18:prl}. Recent measurements with two highly charged lithiumlike calcium isotopes \cite{koehler:16:nc} have demonstrated the possibility to access bound-state QED beyond the Furry picture in the strong coupling regime, specifically the relativistic nuclear recoil effect \cite{shabaev:17:prl,malyshev:17:jetpl,shabaev:18:pra}. While hydrogenlike ions, due to their simplicity, allow for the most accurate theoretical predictions, nuclear effects set the ultimate limits of the theoretical accuracy regardless of the progress in QED calculations. However, in combination with measurements on lithiumlike and boronlike ions, these limits can be overcome \cite{shabaev:02:pra,shabaev:06:prl}. Here, specific differences of the \emph{g}-factor values of different charge states with the same nucleus exhibit orders-of-magnitude smaller theoretical uncertainty than the individual \emph{g} factors \cite{shabaev:02:pra,shabaev:06:prl,volotka:14:prl-np,yerokhin:16:prl}. Based on this, an independent determination of the fine structure constant from heavy hydrogen- and boronlike ions \cite{shabaev:06:prl} and from light hydrogen- and lithiumlike ions \cite{yerokhin:16:prl} has been proposed. Following the experiments with hydrogenlike ions \cite{haeffner:00:prl,verdu:04:prl,sturm:11:prl,sturm:13:pra}, the \emph{g} factor of lithiumlike silicon has been measured at the Mainz University with a relative uncertainty of $1.1\times 10^{-9}$ \cite{wagner:13:prl}. Shortly after, the \emph{g} factors of two lithiumlike calcium isotopes have been measured with two-times smaller uncertainty \cite{koehler:16:nc}. The corresponding efforts devoted to the evaluation of the many-electron contributions to the \emph{g} factor of three-electron ions have led to a theoretical uncertainty of $6\times 10^{-9}$ for silicon \cite{yerokhin:17:pra-2} and $13\times 10^{-9}$ for calcium \cite{volotka:14:prl}. 

In this Letter, we present simultaneous experimental (by a factor of 15) and theoretical (by a factor of 2) improvements of the \emph{g} factor of lithiumlike silicon. In view of the determination of the fine structure constant \cite{yerokhin:16:prl} this represents an important step towards this long-term goal. Meanwhile, the theoretical uncertainty is now dominated by the contributions of the next-order many-electron QED diagrams. In order to achieve further theoretical progress, these contributions need to be evaluated rigorously (to all orders in $\aZ$), while the remaining theoretical background is sufficiently developed to match the present experimental accuracy. 
%
%
%

\textit{Experiment.} ---
The Zeeman splitting of the electron energy levels in a homogeneous magnetic field $B$: $\Delta E=h \nu_L=h {g}{e}/({4\pi}{m_e})B$ gives experimental access to the bound-electron \emph{g} factor. Here, $h$ denotes the Planck constant, $\nu_L$ the Larmor frequency, $e$ the electric charge of the electron and $m_e$ its mass. By measuring the cyclotron frequency of the highly-charged ion $\nu_c=q_\text{ion} B/(2\pi m_\text{ion})$, where $q_\text{ion}$ is the electric charge and $m_\text{ion}$ is the mass of the ion, the magnetic field can be determined and the \emph{g} factor is given by the ratios of frequencies ($\Gamma \equiv \nu_L/\nu_c$), masses ($m_e/m_\text{ion}$), and charges ($q_\text{ion}/e$):
\begin{equation}
  g = 2\,\frac{\nu_L}{\nu_{c}}\,\frac{m_e}{m_\text{ion}}\frac{q_\text{ion}}{e} 
    \equiv 2\Gamma\,\frac{m_{e}}{m_\text{ion}}\,\frac{q_\text{ion}}{e}
\,.
\end{equation}
In case of lithiumlike silicon, the mass of the ion $m({}^{28}\textnormal{Si}^{11+})=27.970\,894\,575\,55 (75)$~u \cite{AME_2016,NIST_IP,CODATA14} and the mass of the electron $m_{e}=0.000\,548\,579\,909\,070 (16)$~u \cite{sturm:14:n,CODATA14,koehler:15:jpb} contribute to the relative systematic uncertainty of the experimentally determined \emph{g} factor on a level of $\delta g/g\mid_{m_\text{ion},m_e} = 4\times 10^{-11}$, whereas the previously measured frequency ratio $\Gamma$ entails a relative uncertainty of $\delta g/g\mid_{\Gamma}=1\times 10^{-9}$ \cite{wagner:13:prl}. In the following, a 15-fold improved value of $\Gamma$ is presented.

For the determination of the Larmor-to-cyclotron frequency ratio $\Gamma$, the experimental apparatus for bound-electron \emph{g} factors of highly charged ions, located in Mainz, has been used \cite{koehler:15:jpb}. Here, single highly-charged ions are trapped and studied in a Penning-trap setup, which is placed in a hermetically sealed, cryogenic vacuum vessel permeated by a homogeneous $3.8$~T magnetic field. In a five electrode cylindrical Penning trap with a radius $r=3.5$~mm and a ring voltage of about $-7.2$~V the three eigenfrequencies, the modified cyclotron frequency $\nu_+\approx22.7$~{MHz}, the axial frequency $\nu_z\approx631$~{kHz} and the magnetron frequency $\nu_-\approx8.8$~{kHz} are measured non-destructively at cryogenic temperatures ($T_z\approx4$~{K}) via a superconductive axial tank circuit. The invariance theorem $\nu_c=\sqrt{\nu_+^2+\nu_z^2+\nu_-^2}$ \cite{brown:86:rmp} is used to determine the free cyclotron frequency. In each measurement cycle a microwave field at randomly chosen, fixed frequency $\nu_L^*$ around the estimated Larmor frequency $\nu_L\approx105.4$~{GHz} is injected for 5 seconds. During this time the modified cyclotron frequency is measured via the phase-sensitive technique PnA \cite{sturm:11:prl-2}. To determine the spin-state of the bound electron by the so-called continuous Stern-Gerlach effect \cite{dehmelt:86:pnas} before and after these measurements, the ion is adiabatically transported from the precision trap (PT) into an adjacent trap, the analysis trap (AT). Here, a large magnetic bottle ($B_2=10500$~{T/m}$^2$) couples the spin-state of the bound electron to the axial motion ($\Delta \nu_z=\nu_z(\uparrow)-\nu_z(\downarrow)\approx260$~{mHz} at $\nu_z\approx 408$~{kHz}). After each measurement cycle, a frequency ratio $\Gamma^*=\nu_L^*/\nu_c$ and the information whether or not the spin has flipped is recorded. Repeating this measurement cycle at different $\nu_L^*$ allows to apply a maximum likelihood fit to the spin-flip probability versus the measured frequency ratios $\Gamma^*$. From this Gaussian $\Gamma$-resonance the frequency ratio $\Gamma_\text{stat}$ is extracted. More details on the different frequency detection techniques, the measurement cycle and the lineshape of the $\Gamma$ resonance can be found, e.g., in Ref.~\cite{koehler:15:jpb}.

In comparison to the previous measurement of $^{28}$\textnormal{Si}$^{11+}$ \cite{wagner:13:prl}, two major improvements have been made. (1) At first, the spin-flip rate in the AT has been increased from $1\%$ to more than $40\%$ with two measures: (a) We increased the microwave power by optimizing the wave guides and their horn-horn transitions. (b) Due to the strong magnetic bottle, the magnetic field and thus the Larmor frequency in the AT is frequency modulated: $B(t) \approx B_0 + B_2\,\delta z^2 + 2 B_2\,\delta z\,z(t)$, where $z(t)=z_0' \sin(\omega_z t + \phi_0)$ is the axial amplitude and $\delta z$ is a shift of the axial center position of the ion with respect to the center position of the magnetic bottle, e.g., caused by intrinsic patch potentials on the electrode surfaces. In case of a non-vanishing modulation index $\eta \equiv \Delta \nu_L/\nu_z$, a sideband structure appears and the spin-flip rate is reduced. Since the modulation index scales linearly with $z_0$, the modulation has been decreased and thus the spin-flip rate increased, by placing the ion in the center of the magnetic bottle. In this way, the measurement cycle time could be reduced by a factor of 5 to 30 minutes. (2) Secondly, the application of the phase sensitive measurement technique PnA \cite{sturm:11:prl-2} instead of the double-dip technique \cite{wagner:13:prl} reduced the measurement time of the modified cyclotron frequency by a factor of 30. 

%
\begin{figure}
\includegraphics[width=0.5\textwidth]{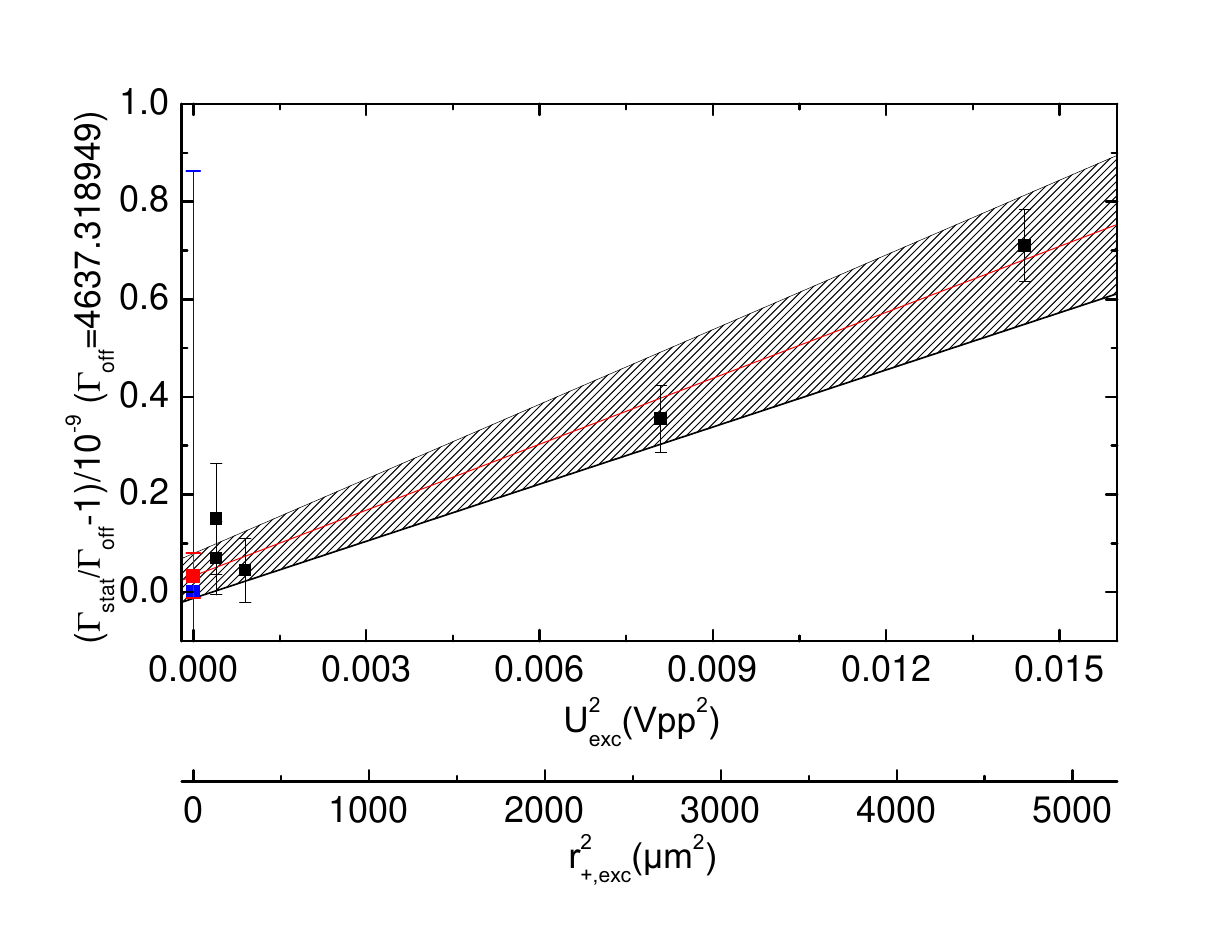}%
\caption{Measured frequency ratios $\Gamma_\text{stat}$ at different modified cyclotron excitation energies during the PnA cycle minus some offset parameter $\Gamma_\text{off}$. The red line indicates a linear fit extrapolating to zero excitation energy. The extrapolated value (red dot) is in excellent agreement with the former measurement (blue dot) \cite{wagner:13:prl}. The grey hatched area indicates the uncertainty of the linear fit.
\label{GammaPlot}}
\end{figure}
\begin{table}
\begin{center}
\caption{Relative systematic shifts of the frequency ratio $\Gamma$ and their uncertainties, defined as $(\Gamma_\text{final}-\Gamma_\text{stat})/\Gamma_\text{off}$.
\label{Sysshifts}}
\begin{ruledtabular}
\begin{tabular}{l c c}
& Rel. shift & Rel. uncert. \\
Effect & ($\times 10^{12}$) & ($\times 10^{12}$) \\
\hline
Image charge                          &  $-659$ &  $33$ \\
Lineshape model                       &     $0$ &   $7$ \\
Residual electrostatic anharmonicity  & $\ll 1$ &   $2$ \\
Magnetron frequency uncertainty       &     $0$ &   $2$ \\
Image current                         &    $-1$ &   $1$ \\
Residual magnetostatic inhomogeneity  &   $0.5$ & $0.4$ \\
Residual special relativity           &  $-0.2$ & $0.3$ \\
\hline
Total                                 &  $-660$ &  $34$ \\
\end{tabular}
\end{ruledtabular}
\end{center}
\end{table}
In 1.5 months 1674 measurement cycles (263 cycles with a spin-flip in the PT and 1411 cycles without a spin-flip in the PT) have been recorded at four different modified cyclotron excitation radii $r_{+,\text{exc}}=$11, 17, 52, 69~$\mu${m}, see Fig.~\ref{GammaPlot}. Here, the slope is mainly given by the relativistic shift of the modified cyclotron energy. The linearly extrapolated frequency ratio at zero excitation energy $\Gamma_{\text{stat}}(r_{+,\text{exc}}=0) = 4637.318\,949\,16\,(21)$ has to be corrected for systematic shifts, see Table~\ref{Sysshifts}, which are also discussed in \cite{koehler:15:jpb}:
\begin{equation}
  \Gamma_\text{final}\left(^{28}\textnormal{Si}^{11+}\right) = 4637.318\,946\,10\,(27)
\,.
\end{equation}
The additional electric field, which is generated by the induced image charges on the trap surfaces, causes small shifts in the two radial frequencies \cite{sturm:13:pra}. This so-called image charge shift is the dominant systematic effect. In combination with the masses given above we determine an improved value of the bound-electron \emph{g} factor of lithiumlike silicon:
\begin{equation}
  g_\text{exp}\left(^{28}\textnormal{Si}^{11+}\right) = 2.000\,889\,888\,450\,(92)\,(68)\,(53)\,(58).
\end{equation}
Here, the statistical and systematic uncertainty of $\Gamma$ as well as the uncertainties due to the mass of $^{28}$\textnormal{Si}$^{11+}$ and the mass of the electron are given in the four brackets separately. 
Our new value is in excellent agreement with the former measurement and exceeds its precision by a factor of 15. With a relative uncertainty of $7.0 \times 10^{-11}$, this new value also surpasses the precision of the currently most accurate lithiumlike \emph{g} factor ($\delta g_\text{exp}\left(^{48}\textnormal{Ca}^{17+}\right) / g = 4.1 \times 10^{-10}$ \cite{koehler:16:nc}) by a factor of 6.
%
%
%

\textit{Theory.} ---
While for one-electron systems the theoretical consideration of the \emph{g} factor comes down to the QED and nuclear effects, for lithiumlike ions the electron-electron interaction effects come into play. In contrast to other atomic properties, such as binding energies, for the \emph{g} factor these effects are purely relativistic, i.e., they vanish in the non-relativistic limit. Moreover, the contribution of the negative-energy states is not suppressed as compared to the positive-energy states and is equally important. These features make the \emph{g}-factor evaluation in many-electron systems more involved than, e.g., the evaluation of the binding energies. Various calculation methods have been employed over the years, which resulted in today's accuracy on the level of $10^{-9}$. In general, there are three expansion parameters that can be used in the theoretical description of highly charged ions: $\alpha$, $\aZ$, and $1/Z$, where $\alpha$ is the fine structure constant and $Z$ is the nuclear charge number. Different theoretical approaches rely on the expansions in $\aZ$, $1/Z$, or both. The rigorous QED approach accounts for all orders in $\aZ$, while only few leading orders in $\alpha$ and $1/Z$ are accessible up-to-date. In particular, the diagrams of the one- and two-photon exchange ($\sim 1/Z$ and $\sim 1/Z^2$) and the two-electron self-energy and vacuum-polarization diagrams ($\sim \alpha/Z$) have been evaluated to all orders in $\aZ$ \cite{volotka:09:prl,glazov:10:pra,wagner:13:prl,volotka:14:prl}. In turn, the so-called NRQED (nonrelativistic quantum electrodynamics) approach provides access to all orders in $1/Z$ (based on the Schr\"odinger equation), however, it is restricted to the leading orders in $\aZ$ (see, e.g., \cite{wienczek:14:pra,yerokhin:17:pra-2}). The interelectronic-interaction operator incorporating the leading relativistic corrections is known as the Coulomb-Breit operator. For this reason, the term ``Breit approximation'' is widely used for the results obtained with this operator and more generally for the results, which reproduce correctly the contributions to the \emph{g} factor of the order $(\aZ)^2$. The higher-order remainder (starting from $(\aZ)^4$) can be termed as a ``non-trivial QED contribution''. However, this separation depends on the exact formulation of the approach used to obtain the Breit approximation.

The interelectronic-interaction effects determine the accuracy of the recently published theoretical \emph{g}-factor values~\cite{volotka:14:prl,yerokhin:17:pra-2} in the middle-$Z$ region. For this reason, below we consider these effects in some detail. According to the previous works, we separate the many-electron contributions into the pure interelectronic-interaction correction, the screening of the QED effects, and the corresponding correction to the nuclear recoil effect. The effect of the finite nuclear size can be taken into account for each of these contributions by using the proper nuclear Coulomb potential. 

First, we consider the interelectronic-interaction correction to the \emph{g} factor of lithiumlike ions within the Breit approximation. Accurate calculations to all orders in $1/Z$ have been performed by Yan \cite{yan:01:prl,yan:02:jpb} based on the effective two-component Hamiltonian derived by Hegstrom \cite{hegstrom:73:pra,hegstrom:75:pra}. Recently, Yerokhin and co-authors have performed similar calculations with much better accuracy \cite{yerokhin:17:pra-2}. In this work, we present an independent calculation within the recursive formulation of the perturbation theory~\cite{glazov:17:nimb}. This method proved to be efficient for calculations of the higher-order interelectronic-interaction contributions to the binding energies in few-electron ions~\cite{glazov:17:nimb,malyshev:17:pra,malyshev:19:pra} and to the nuclear-recoil effect on the \emph{g} factor~\cite{shabaev:17:prl,shabaev:18:pra}. Extension of this method to the \emph{g}-factor calculations implies proper account for the contribution of the negative-energy continuum. The key advantage of this method is that it provides direct access to the individual terms of the $1/Z$-expansion. Consequently, no fitting procedure is needed to identify the part of the order $1/Z^3$ and higher, which is to be combined with the QED values for the $1/Z$ and $1/Z^2$ terms. The convergence of the $1/Z$-expansion can be improved significantly by using the effective screening potential in the Dirac equation, which defines the zeroth-order wave functions. We consider three different screening potentials --- core-Hartree, Kohn-Sham, and local Dirac-Fock~\cite{kohn:65:pr,sapirstein:02:pra,shabaev:05:pra,glazov:06:pla,malyshev:17:pra}. As a result, we find the Breit-approximation part of the interelectronic interaction with an uncertainty on the level of $1\times 10^{-9}$. It has to be complemented by the non-trivial QED contribution (higher orders in $\aZ$) evaluated with the same screening potential.

Evaluation of the interelectronic interaction to all orders in $\aZ$ within the framework of bound-state QED can be done only order by order in $1/Z$. The first-order correction (one-photon exchange) is relatively simple, it has been calculated, e.g., in Ref.~\cite{shabaev:02:pra} for a wide range of $Z$. The two-photon-exchange correction is significantly more involved, including the derivation of the complete set of formulas and development of the numerical procedure. The first evaluation for lithiumlike silicon with the Coulomb potential in Ref.~\cite{wagner:13:prl} was extended to several other lithiumlike ions and to various screening potentials in Ref.~\cite{volotka:14:prl}. In this paper, we reevaluate the one- and two-photon-exchange contributions for silicon with the potentials listed above to match the Breit-approximation values. The total interelectronic-interaction contribution to the \emph{g} factor of lithiumlike silicon is $314.8118\,(12)\,(24)\times 10^{-6}$. The first error bar here is the numerical uncertainty of the calculations. The second one is due to the unknown nontrivial QED contribution of the three-photon-exchange diagrams. It can be estimated based on different ratios of the presently known contributions. As an approximate average of these estimations, we use the expression $2\,(\aZ)^2\,\Delta g^{(3)}$, where $\Delta g^{(3)}$ is the $1/Z^3$-contribution evaluated in the Breit approximation. This estimation is more conservative than the one used in Ref.~\cite{yerokhin:17:pra-2}. 

The interplay between the interelectronic-interaction and QED effects leads to the two-electron or ``screened'' QED correction. In analogy to the ``pure'' interelectronic-interaction contribution considered above, one can also consider the Breit approximation here. To this end, one can use the set of two-component effective QED operators \cite{hegstrom:73:pra,hegstrom:75:pra}. For $s$-states these operators yield the correct result up to the order $(\aZ)^2$ for arbitrary order in $\alpha$ and $1/Z$. In Refs.~\cite{yan:01:prl,yan:02:jpb,yerokhin:17:pra-2} these operators were used to evaluate the screened QED correction by averaging with the many-electron wave functions obtained from the many-electron Schr\"odinger equation. In Ref.~\cite{glazov:04:pra} the four-component counterparts of these operators were used to calculate the $1/Z$ contribution. In this work, we incorporate these operators in the recursive perturbation theory in order to find the contributions of arbitrary order in $1/Z$. In order to obtain the sought-for contributions we develop the multi-recursive scheme of the perturbation theory with respect to the following operators: the effective four-component QED operators (first order), the magnetic-field interaction (first order), and the interelectronic interaction (arbitrary order). In addition, we employ the effective screening potential (see above), which accelerates the convergence of the perturbation series in $1/Z$. 

Screened QED correction of the first orders in $\alpha$ and in $1/Z$ corresponds to the set of two-electron self-energy and vacuum-polarization diagrams, which have been evaluated to all orders in $\aZ$ in Refs.~\cite{volotka:09:prl,glazov:10:pra,volotka:14:prl}. The numerical uncertainty of these calculations gets larger for smaller nuclear charge due to the large cancellations of individual terms and the poor partial-wave convergence. In this work, in order to get the most of both the rigorous approach and the effective operators, we have performed the calculations within both methods for a set of $Z$ in the range 20--50. Then the difference between these values has been extrapolated to $Z=14$ by fitting to the polynomials in $1/Z$ and $\aZ$. As a result, we find $-0.2415\,(14)\,(16)\times 10^{-6}$ for the screened QED correction to the \emph{g} factor of lithiumlike silicon. The first given uncertainty is numerical and the second one is due to the unknown nontrivial QED contribution of the second and higher orders in $1/Z$. It is estimated using the ratio of the nontrivial QED contribution and the Breit-approximation part of the $1/Z$ contribution, in full agreement with Ref.~\cite{yerokhin:17:pra-2}. 

The theoretical results for the \emph{g} factor of lithiumlike silicon are summarized in Table~\ref{tab:total-g}. The finite nuclear size effect is calculated numerically, the uncertainties due to the nuclear radius and model are negligible at present. The interelectronic-interaction and the screened QED contributions are evaluated in the present work as described above. The one-loop one-electron QED correction is taken from Refs.~\cite{glazov:04:pra,lee:05:pra,yerokhin:17:pra-1}. The QED correction of the second and higher orders in $\alpha$ obtained within the framework of $\aZ$-expansion is taken from Refs.~\cite{pachucki:05:pra,yerokhin:17:pra-2,czarnecki:16:pra,czarnecki:18:prl}. For the nuclear recoil effect we use the most recent results from Refs.~\cite{shabaev:17:prl,shabaev:18:pra}, which include the higher-order terms in $\aZ$ and the interelectronic-interaction contributions. We note that the value of the nuclear recoil correction has changed considerably due to the many-electron part, which was found to be treated incompletely in previous works. The total theoretical value of the \emph{g} factor is 
\begin{equation}
  g_\text{th}\left(^{28}\textnormal{Si}^{11+}\right) = 2.000\,889\,894\,4\,(34)
\,.
\end{equation}
The error bar is largely determined by the estimation of the presently unknown contributions of the two-loop many-electron diagrams: three-photon exchange and two-photon exchange with additional self-energy loop. The difference between $g_\text{th}$ and $g_\text{exp}$ is $1.7$ times larger than this uncertainty which strongly stimulates further theoretical investigations.
\begin{table}
\caption{Individual contributions to the ground-state \emph{g} factor of lithiumlike silicon and comparison with the experimental result and with the previously reported theoretical and experimental values. The experimental result of Wagner {\it et al.} \cite{wagner:13:prl} is updated for the new values of the electron mass and the mass of $^{28}$\textnormal{Si}$^{11+}$.
\label{tab:total-g}}
\vspace{0.25cm}
\begin{tabular}{lr@{}l} 
\hline
\hline
%
Dirac value (point nucleus) &   1.&998\,254\,750\,7        \\
Finite nuclear size         &   0.&000\,000\,002\,6        \\
QED, $\sim \alpha$          &   0.&002\,324\,043\,9        \\
QED, $\sim \alpha^{2+}$     &$-$0.&000\,003\,516\,6\,(3)   \\
Interelectronic interaction &   0.&000\,314\,811\,8\,(27)  \\
Screened QED                &$-$0.&000\,000\,241\,5\,(21)  \\
Nuclear recoil              &   0.&000\,000\,043\,6        \\
\hline
Total theory, this work     &   2.&000\,889\,894\,4\,(34)  \\
Total theory, Yerokhin {\it et al.} \cite{yerokhin:17:pra-2}
                            &   2.&000\,889\,892\,(6)      \\
Total theory, Volotka {\it et al.} \cite{volotka:14:prl}
                            &   2.&000\,889\,892\,(8)      \\
\hline
Experiment, this work       &   2.&000\,889\,888\,458\,(137) \\
Experiment, Wagner {\it et al.} \cite{wagner:13:prl}
                            &   2.&000\,889\,888\,4\,(19)$^*$\\
\hline\hline
\end{tabular}
\\
\vspace{0.25cm}
$^*$updated for the involved mass values (see text).
\end{table}
%

%
%
\textit{Conclusion.} ---
In summary, we have presented a 15-fold improvement of the experimental value and a 2-fold improvement of the theoretical value of the \emph{g} factor of $^{28}$\textnormal{Si}$^{11+}$. The experimental and theoretical relative uncertainties amount to $7.0\times 10^{-11}$ and $1.7\times 10^{-9}$, respectively. The latter is mostly determined by the unknown many-electron two-loop QED contributions. The obtained values are $1.7\,\sigma$ apart, which may indicate that these contributions exceed our present estimations. Further laborious developments of the theoretical methods are required to resolve this discrepancy. At the same time, the obtained experimental value has a potential to validate the non-trivial parts of the many-electron two-loop QED contributions on a few percent level.
%

We want to thank Sascha Rau for fruitful discussions during the final analysis. This work was supported by RFBR (Grants No. 16-02-00334 and 19-02-00974), by DFG (Grant No. VO 1707/1-3), by SPbSU-DFG (Grants No. 11.65.41.2017 and No. STO 346/5-1), by the Max Planck Society, and by the International Max Planck Research School for Quantum Dynamics in Physics, Chemistry and Biology (IMPRS-QD).
%
%

%
%
\end{document}